\documentclass[12pt,a4]{iopart} 
\usepackage{colordvi,epsfig} 
\usepackage{pstricks}
\usepackage{psfrag}
\usepackage{pst-all}
\usepackage{subfigure}
\usepackage{amssymb}
\usepackage{bm}
\usepackage{cite}

\begin{document}

\title[3D Equilibrium and stability for MAST]{Modelling of three dimensional equilibrium and stability of MAST plasmas with magnetic perturbations using VMEC and COBRA}
\author{C J Ham, I T Chapman, A Kirk, S Saarelma}
\address{EURATOM/CCFE Fusion Association, Culham Science Centre, Abingdon, Oxon, OX14 3DB, UK.}
\ead{christopher.ham@ccfe.ac.uk}
\date{\today}




\begin{abstract}
It is known that magnetic perturbations can mitigate edge localized modes (ELMs) in experiments, for example MAST (Kirk {\it et al} 2013 {\it Nucl. Fusion} {\bf 53} 043007 ). One hypothesis is that the magnetic perturbations cause a three dimensional corrugation of the plasma and this corrugated plasma has different stability properties to peeling-ballooning modes compared to an axisymmetric plasma. It has been shown in an up-down symmetric plasma that magnetic perturbations in tokamaks will break the usual axisymmetry of the plasma causing three dimensional displacements (Chapman {\it et al} 2012 {\it Plasma Phys. Control. Fusion} {\bf 54} 105013). We produce a free boundary three-dimensional equilibrium of a lower single null MAST relevant plasma using VMEC (S P Hirshman and J C Whitson 1983 {\it Phys. Fluids} {\bf 26} 3553). The current and pressure profiles used for the modelling are similar to those deduced from axisymmetric analysis of experimental data with ELMs. We focus on the effect of applying $n=3$ and $n=6$ magnetic perturbations using the RMP coils. A midplane displacement of over $\pm 1$ cm is seen when the full current is applied. The current in the coils is scaned and a linear relationship between coil current and midplane displacement is found. The effect of this non-axisymmetric equilibrium on infinite $n$ ballooning stability is investigated using COBRA ( R Sanchez {\it et al} 2000 {\it J. Comput. Phys.} {\bf 161} 576-588). The infinite $n$ ballooning stability is analysed for two reasons; it may give an indication of the effect of non-axisymmetry on finite $n$ peeling-ballooning modes, responsible for ELMs; and infinite $n$ ballooning modes are correlated to kinetic ballooning modes (KBMs) which are thought to limit the pressure gradient of the pedestal (Snyder {\it et al} 2009 {\it Phys. Plasmas} {\bf 16} 056118). The equilibria with midplane displacements due to RMP coils have a higher ballooning mode growth rate than the axisymmetric case and the possible implications are discussed.   
\end{abstract}
\submitto{Nuclear Fusion}
\maketitle

\section{Introduction} 
\label{sec:Intro}

It is well known that edge localized modes (ELMs) need to be controlled in ITER because they may limit the lifetime of the machine \cite{Loarte03}. There are several possible strategies for controlling ELMs such as pellets or vertical kicks but we focus on in vessel resonant magnetic perturbation (RMP) coils here which produce a non-axisymmetric magnetic perturbation to the plasma. It has been demonstrated in many machines that this can either mitigate (i.e. increase the frequency of ELMs and reduce the peak heat flux) or suppress (i.e. remove ELMs completely) ELMs depending on toroidal mode number, parity and target plasma \cite{Kirk13}. In MAST ELMs have been mitigated in connected double null configuration (CDN) with $n=3$ and in lower single null (SND) with $n=4$ and $n=6$ \cite{Kirk13}.

The magnetic perturbations are not axisymmetric and it has been shown on MAST that when the magnetic perturbations are applied the plasma gains a non-axisymmetric, three dimensional character. Non-axisymmetry has been observed experimentally in MAST where the outboard miplane gains a corrugation \cite{Chapman12} and also at the X-point where the plasma forms lobe structures which are characteristic of the homoclinic tangle \cite{Kirk13}. The effect of non-axisymmetric fields has been seen experimentally or modelled for several machines, see for example \cite{Chapman07, Canik12, Fischer12, Chapman13} . 

The stability of the non-axisymmetric equilibria to infinite $n$ ballooning modes is of interest for two reasons. First, although the ELM is thought to be linked to finite $n$ peeling-ballooning modes the infinite $n$ results should give an indication of the effect of non-axisymmetry on the stability of finite $n$ modes. The calculation of finite $n$ stability in stellarator geometry is quite numerically heavy especially for tight aspect ratio experiments, such as MAST. Secondly, the stability of infinite $n$ ballooning modes correlates well with the stability of the kinetic ballooning modes (KBMs) which are thought to limit the gradient of the edge pedestal \cite{Snyder09, Dickinson11}.     

The analysis of equilibrium and stability with the magnetic perturbations applied has until recently been based on axisymmetric models. However, many of the assumptions behind these models rely on symmetry arguments and these models may therefore not capture the full physics. It fact axisymmetric intuition may hinder our understanding. We must understand if three dimensional effects are important in ELM mitigation or suppression with RMPs. Non-axisymmetry is a much more difficult problem than axisymmetry and so non-axisymmetric models should only be used if they are necessary.

There are a number of approches to studying non-axisymmetric plasmas as discussed in \cite{Turnbull13}. The approaches can be categorized as either linear or nonlinear and either dynamic evolution or nearby equilibrium. Linear, dynamic evolution codes generally take an axisymmetric equilibrium with a non-axisymmetric perturbation to find the linearly perturbed state, codes include MARS-F \cite{Liu00}. Quasi-linear terms such as torques on the plasma can be included, as in MARS-Q \cite{Liu12}. Nonlinear, dynamic evolution codes seek to find saturated states of the equations, examples include M3D-C$^1$ \cite{Ferraro10} and NIMROD \cite{Glasser99}. However, these codes have a significant run time possibly requiring tens of thousands of CPU-hours for a single result \cite{Turnbull13}. The nearby equilibrium approach can be linear such as IPEC \cite{Park07} or nonlinear, which is the approach we use here. There are several codes, developed by the stellarator community, which look at this problem. We will use the VMEC here \cite{Hirshman83, Hirshman86} which assumes nested flux surfaces. Codes such as SPEC \cite{Hudson12}, SIESTA \cite{Hirshman11} or HINT2 \cite{Suzuki06} do not make this assumption and so magnetic islands can form in these codes.   

In section~\ref{sec:Equilibrium} we will use free boundary VMEC \cite{Hirshman83, Hirshman86} to calculate a non-axisymmetric equilibrium for a MAST relevant case with and without RMPs. We then impose magnetic perturbations with scans in current for the  toroidal mode number $n=3$ case. The stability of these equilibria is investigated in section~\ref{sec:Stability} using the COBRA code \cite{Sanchez00} which calculates the infinite $n$ ballooning stability of the equilibrium. We will discuss the results and give conclusions in section~\ref{sec:Conclusion}.  

\section{Equilibrium}
\label{sec:Equilibrium}

We calculate the plasma equilibrium using the free boundary VMEC and the plasma poloidal and toroidal field coils, assuming no error field. We then apply RMPs to this equilibrium to find the new non-axisymmetric equilibrium. 

\subsection{VMEC}

Axisymmetric plasma equilibrium can be reduced to solving the Grad-Shafranov equation, which is a partial differential equation, with given pressure and current profiles. The loss of axisymmetry in general means that there is no longer such a partial differential equation. The approach used by VMEC to find plasma equilibria is minimization of the total plasma energy
\begin{equation}
W=\int \left(\frac{B^2}{2\mu_0}+\frac{p}{\gamma-1}\right)\textup{d}^3x,
\end{equation}
where $p$ is the plasma pressure, $B$ is the magnetic field, and $\gamma$ is the adiabatic index. VMEC does not allow solutions that contain magnetic islands or ergodic regions, only nested flux surfaces. It may be that ergodic regions or magnetic islands are important in the physics of ELM control with magnetic perturbations however we do not investigate these effects here.

VMEC does not include the effects of rotation or any shielding of the magnetic perturbation. The perturbation is assumed to be fully penetrated. However, VMEC is solving the non-linear equilibrium problem. 
  
\subsection{Axisymmetric equilibrium}

\begin{figure}
\centering
\includegraphics[width=0.8\textwidth]{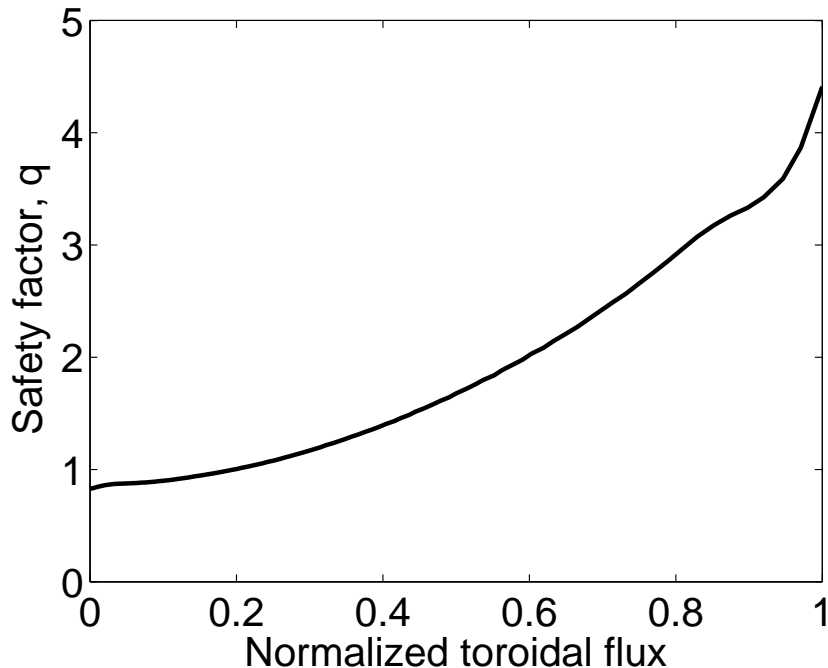}
\caption{Safety factor profile used in VMEC for the cases with and without RMP coils.}
\label{fig:Qprof}
\end{figure}

\begin{figure}
\centering
\includegraphics[width=0.8\textwidth]{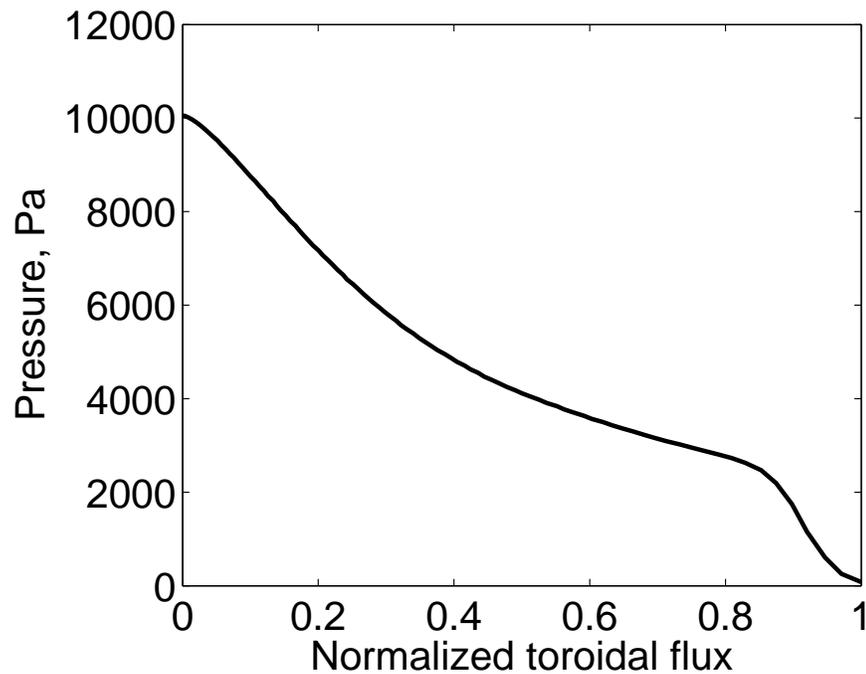}
\caption{Pressure profile used in VMEC. This is a representative pressure profile for a LSND plasma in MAST.}
\label{fig:Pprof}
\end{figure}

\begin{figure}
\centering
\includegraphics[width=0.8\textwidth]{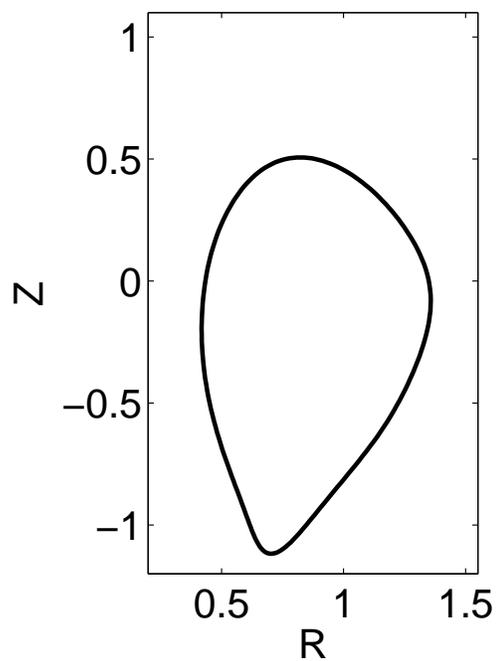}
\caption{A poloidal section of the plasma boundary computed using the pressure and safety factor profiles above.}
\label{fig:Bnd}
\end{figure}

The aim of this paper is to show the effect of applying non-axisymmetric fields to tokamak plasmas. The pressure and safety factor profiles used here are shown in Figures \ref{fig:Qprof} and \ref{fig:Pprof}. They are realistic profiles typical of an axisymmetric reconstruction using EFIT, which includes constraining the reconstructed equilibrium with magnetics, Thomson Scattering, motional Stark effect and $D_{\alpha}$ light at the boundary \cite{Debock12}, for a lower single null (LSND) plasma. The plasma boundary for this equilibrium is shown in Figure \ref{fig:Bnd}.  

\subsection{3D Equilibrium}

We now turn on the magnetic perturbation coils and look at the response of the plasma. MAST has an upper and lower row of magnetic perturbation coils \cite{Kirk13}. However, for the LSND plasma the response is dominated by the lower row of coils because they are so much closer to the plasma and the perturbation field falls very quickly with distance from the coils. We focus on the cases where the lower coils have been applied in $n=3$ and $n=6$ configurations. The qualitative results for other configurations are similar.  

\begin{figure}
\centering
\includegraphics[width=0.8\textwidth]{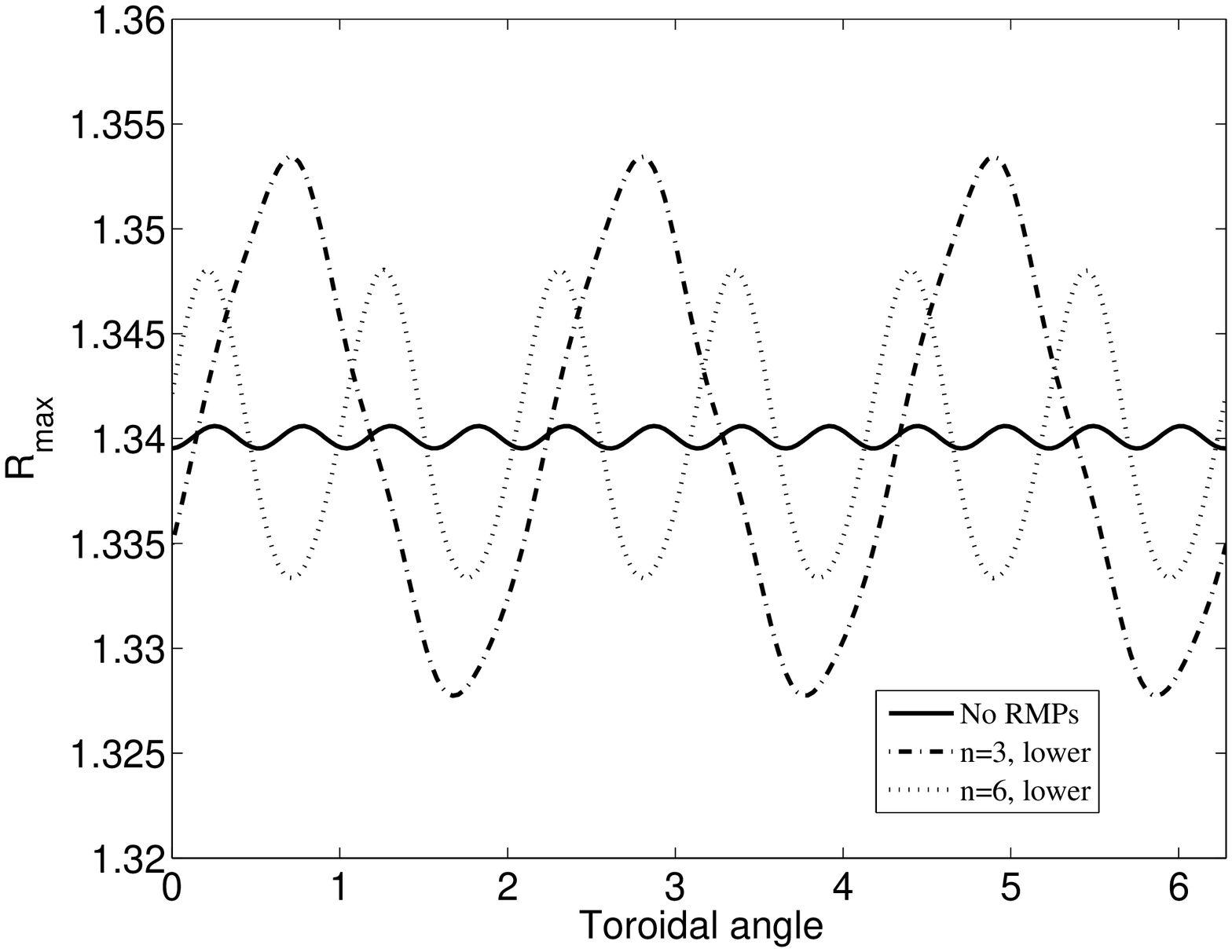}
\caption{Location of the outer flux surface on the magnetic axis plane with RMP coils off (solid); lower RMP coils on $n=3$ (dash-dot); and lower RMP coils on $n=6$ (dotted).}
\label{fig:Rmax}
\end{figure}

Figure \ref{fig:Rmax} shows the location of the outer flux surface located on the magnetic axis plane. The solid line shows the result without the RMP coils switched on. The $n=12$ variation is due to the toroidal field ripple. This ripple is too small to be seen experimentally. The dash-dot line shows the case with the lower RMP coils on in $n=3$ configuration. The plasma gains an $n=3$ toroidal corrugation of over $\pm 1$ cm. The dotted line shows the case with the lower RMPs on in $n=6$ configuration. The corrugation is smaller in amplitude but it is $n=6$ in character.     

\begin{figure}
\centering
\includegraphics[width=0.8\textwidth]{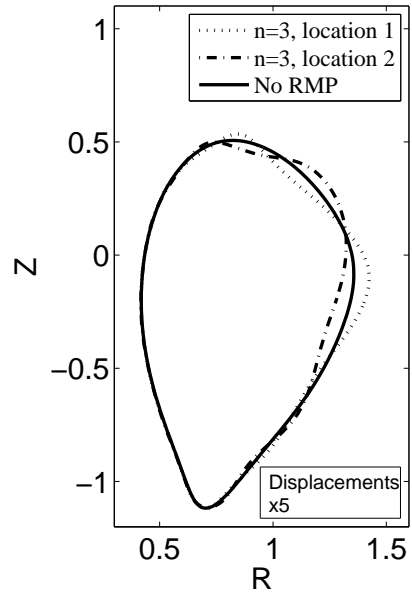}
\caption{A poloidal cut through the plasma. The solid line is the unperturbed case. The dotted and dash-dotted lines are different toroidal locations with the RMP coils on in $n=3$ configuration. The difference between the perturbed and unperturbed cases has been multiplied by five to aid the eye. }
\label{fig:BndL3}
\end{figure}

Figure \ref{fig:BndL3} shows a poloidal section of the plasma without the RMP coils on (solid line) and with the RMP coils on in different toroidal locations (dotted and dash-dotted lines) . The difference between the coils on and coils off case has been multiplied by five to make the difference more visible. The plasma gains helical perturbations especially on the low field side. The high field side is relatively unaffected. The plasma has found a new lower energy state which has this helical structure. The RMP coils have made a `saturated external kink-like' state favourable over the axisymmetric case. 

\subsection{Coil current scans}

\begin{figure}
\centering
\includegraphics[width=0.8\textwidth]{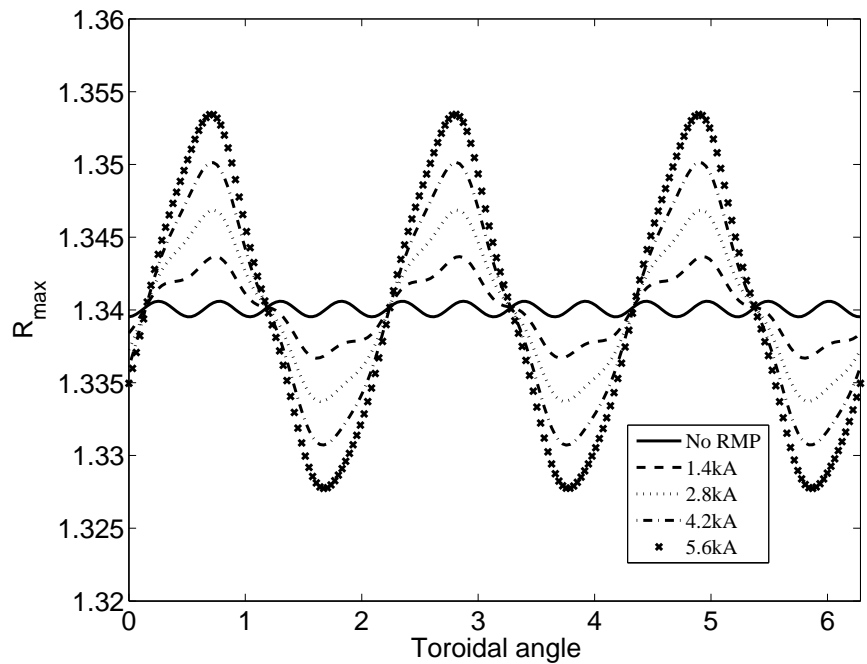}
\caption{Location of the outer flux surface on the magnetic axis plane with RMP coils off (solid) and RMP coils on with 1.4kA (dashed), 2.8kA (dotted), 4.2 kA (dash-dotted) and 5.6kA (x).}
\label{fig:RmaxScan}
\end{figure}

The current in the RMP coils has been increased in steps of a quarter of the maximum, 1.4kA, as shown in figure \ref{fig:RmaxScan}. The corrugation increases linearly with the current in the coils and no threshold current is observed. This is consistent with these corrugations being an ideal MHD effect just arising from the magnetic field perturbation. If resistivity, field penetration and shielding effects are included there would likely be a threshold effect as is seen in the experimental data on the length of the lobes at the X-point \cite{Harrison13}.

\section{Stability}
\label{sec:Stability}

The three dimensional equilibrium alone is not sufficient to understand the effect of RMPs on ELMs. ELMs are thought to be driven by peeling-ballooning modes \cite{Connor98} which are intermediate $n$ modes. However, we do not calculate peeling-ballooning stability of these equilibria here, although this is a future objective. Instead we calculate the infinite $n$ stability to gain a first approximation to the likely effect on the finite $n$ stability.  ELMs are also effected by the edge pressure gradient which is in turn controlled by instabilities at the edge such as the kinetic ballooning mode (KBM). Infinite $n$ stability is correlated to KBM stability. However the growth rate of KBMs is less than the infinite $n$ ballooning modes because of the ion drift resonance and dissipative effects \cite{Snyder09}. We will use the infinite $n$ ballooning mode stability for insight into the stability of peeling-ballooning modes and KBMs here.
  
\subsection{COBRAVMEC}

The stability of these equilibria to infinite $n$ ballooning modes is investigated using COBRA \cite{Sanchez00, Sanchez01}, which uses a fast method to assess stability against ideal infinite $n$ ballooning modes \cite{Connor79, Dewar83}. The ballooning mode growth rate, normalized to the Alfven time, has been calculated on a poloidal and toroidal grid on each flux surface and the maximum value found.

\subsection{Stability and coil configuration}

\begin{figure}
\centering
\includegraphics[width=1.0\textwidth]{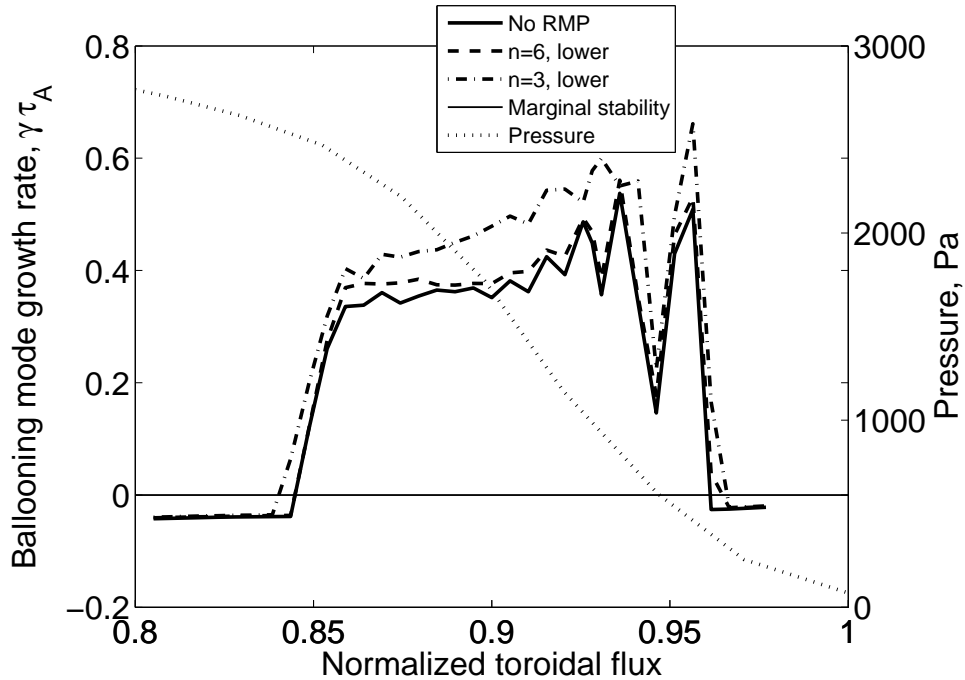}
\caption{The ballooning mode growth rate of the most unstable mode against normalized toroidal flux for the case without RMPs applied (solid line); with RMPs in $n=3$ configuration (dash-dot line); and with RMPs applied in $n=6$ configuration (dashed line). The pressure profile in this region is also plotted (dotted line).}
\label{fig:cobra}
\end{figure}

Figure \ref{fig:cobra} shows the ballooning mode growth rates for the most unstable mode calculated by COBRA for the case with the RMPs applied in $n=3$ (dotted line) and $n=6$ (dot-dash line) and without (solid line). The cases with RMP coils applied has a higher ballooning mode growth rate at all the normalized toroidal flux locations. The $n=3$ case has a higher growth rate because the corrugations are larger for this case.

\begin{figure}
\centering
\includegraphics[width=0.8\textwidth]{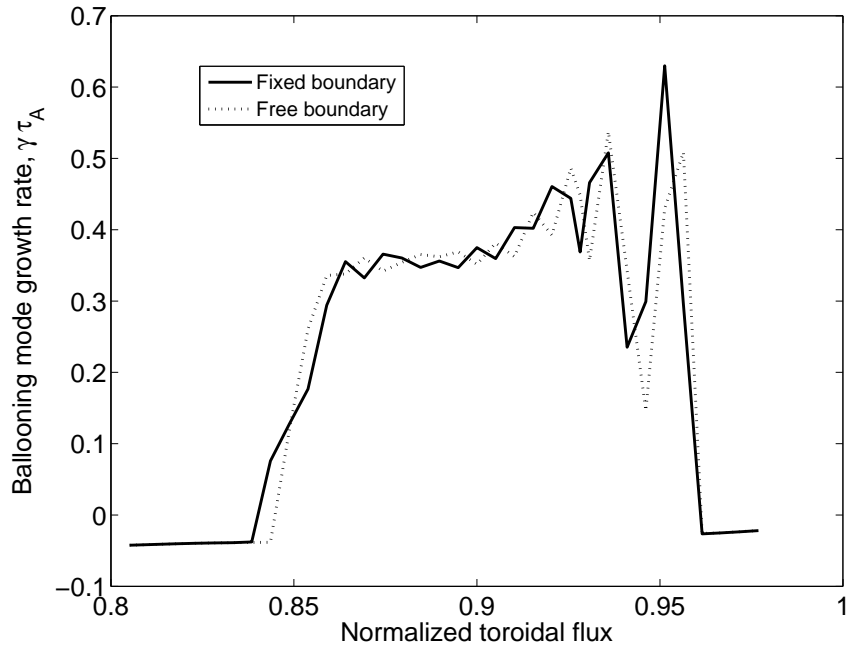}
\caption{The ballooning mode growth rate calculated without RMPs applied with a free boundary (solid line) and a fixed boundary (dashed line). }
\label{fig:cobfixedfree}
\end{figure}

Figure \ref{fig:cobfixedfree} shows that ballooning mode growth rate without the RMPs applied for the free boundary equilibrium and a fixed boundary equilibrium. It is known that an equilibrium produced with a fixed boundary can have slightly different ballooning mode growth rates than a code produced with a free boundary \cite{Sanchez97}. This figure shows that the difference is minimal here. 

\begin{figure}
\centering
\includegraphics[width=0.8\textwidth]{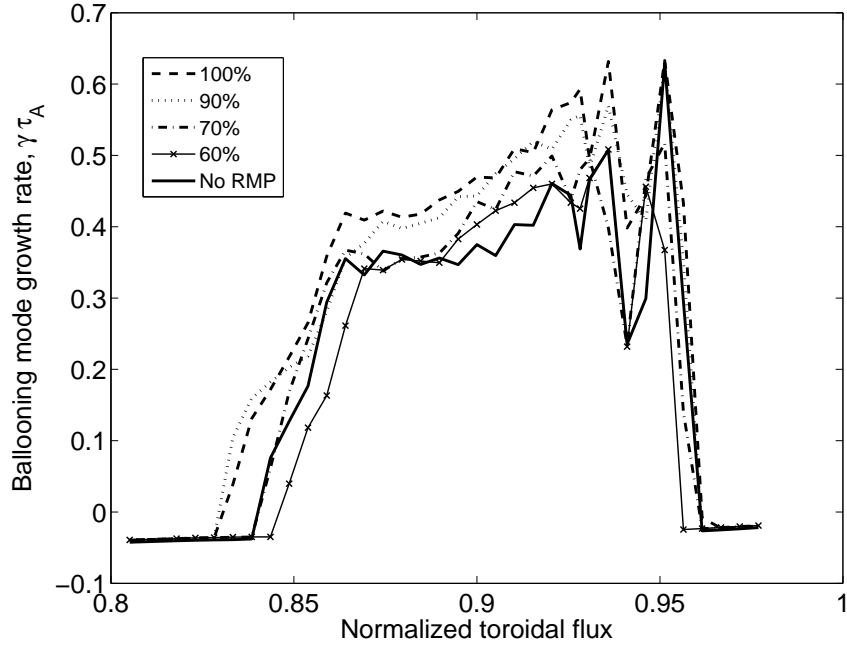}
\caption{The ballooning mode growth rate calculated for fixed boundary equilibria with scales values of the pressure 100\% (dashed); 90\% (dotted); 70 \% (dash-dotted); and 60 \% (-x-). The no RMP case is shown with a solid line.}
\label{fig:cobraFixed}
\end{figure}

It is of interest to calculate the effect of the RMP coils in terms of the pressure change with the RMPs on required to produce the same level of growth and the same marginal points to the case without RMPs applied. If the pressure is scaled in the free boundary calculation the size of the plasma corrugation will change and the safety factor profile may change too, depending on which profiles are kept constant while the pressure profile is changed. These problems have been removed by taking the full pressure, free boundary case and then using the boundary from this as a fixed boundary for new cases with fixed safety factor profile and scaled pressure. Figure \ref{fig:cobraFixed} shows the ballooning mode growth rate with different levels of pressure scaling and fixed boundary. The best match occurs for a pressure scaled to approximately 70\% of the original pressure.

\subsection{Scan of stability with coil current}

\begin{figure}
\centering
\includegraphics[width=0.8\textwidth]{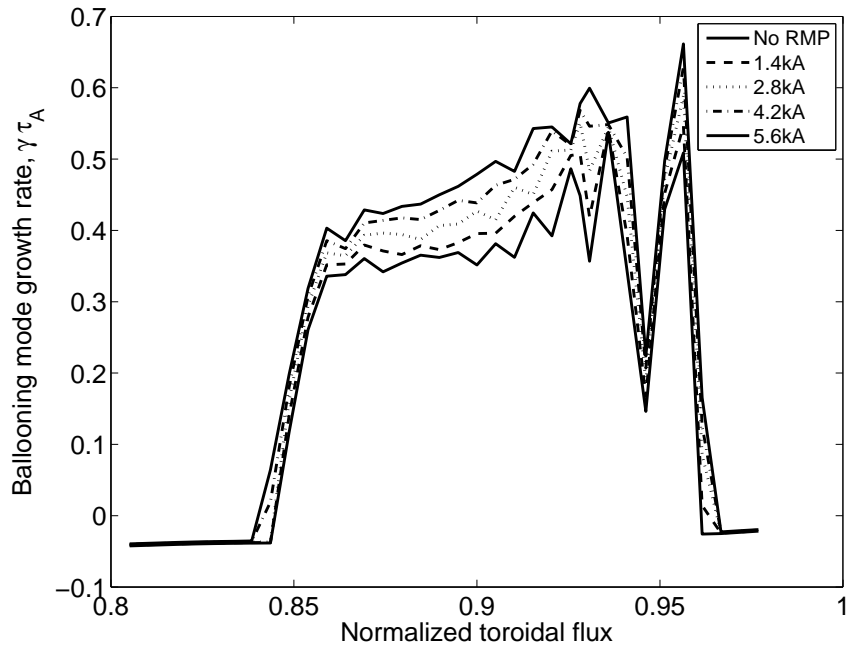}
\caption{The ballooning mode growth rate for free boundary equilibria while the current in the RMP coils is increased 0kA (solid); 1.4kA (dashed); 2.8kA (dotted); 4.2 kA (dash-dotted); and 5.6 kA (-x-).}
\label{fig:cobraIscan}
\end{figure}

Figure \ref{fig:cobraIscan} shows the ballooning mode growth rate for the free boundary equilibria with the current in the RMP coils increased in steps of 1.4kA. The higher the current in the RMP coils the larger the ballooning mode growth rate. The larger RMP coil current produces large corrugations and so a large change in field line curvature, making the most unstable modes more unstable.

\section{Discussion and Conclusions}
\label{sec:Conclusion}

We have calculated free boundary equilibria for an experimentally realistic lower single null plasma using VMEC. We then applied currents in the RMP coils to produce new equilibria and demonstrated that the plasma gains a midplane displacement of over $\pm$ 1cm for the case with the lower RMP coils in $n=3$ configuration. The case with the RMP coils in $n=6$ configuration produced a smaller, but still significant, corrugation. The current in the RMP coils was scanned and a linear relationship between the current in the coils and the corrugation was found. There was no threshold but this is because this is an ideal MHD model which assumes the fields are fully penetrated. 

We then investigated the infinite $n$ stability of these equilibria. We found that the application of RMP coils makes the ballooning modes in the edge region more unstable. There are two possible outcomes from this, as discussed in Chapman {\it et al} \cite{Chapman13b}. First, as the infinite $n$ ballooning stability is linked to the KBM stability, which in turn sets the pedestal gradient, we would expect the pedestal gradient to be lower when RMPs are applied. This has been seen experimentally in MAST \cite{Scannell13}. Secondly, the infinite $n$ ballooning modes may provide insights to the finite $n$ peeling-ballooning modes which are thought to drive the ELMs. Our results indicate that these modes too would become more unstable and so ELMs would become more frequent which is also seen experimentally. The end result of applying RMPs will depend on the relative strengths of these two effects. If the effect on peeling-ballooning modes is dominant then ELMs will become more unstable and more frequent. This may be what happens in ELM mitigation.     

This problem has been investigated analytically by Bird and Hegna \cite{Bird13}. In this work a local equilibrium model was used to assess the effect of three dimensional perturbations on infinite $n$ ballooning stability. The surfaces were only perturbed on the order of a few millimeters but a destabilizing effect was observed. The authors found that the destabilization of infinite $n$ ballooning modes was mostly due to the helical Pfirsch-Schluter currents, which arise when there is a three dimensional perturbation, altering the local shear. In our paper the flux surfaces are moved by a centimeter and so we would expect the change in the curvature of the field lines to be important.

We have used a three dimensional equilibrium model which does not include plasma flow and assumes nested flux surfaces. This has a number of limitations. First, the applied fields are assumed to be fully penetrated into the plasma. However, experimental observations on lobes at the X-points indicate that there is plasma shielding. Second, the application of the RMPs has been observed to slow the plasma rotation and in certain circumstances the plasma stops completely \cite{Kirk13}. The quasi-linear MARS-Q code, for example, is required to model this \cite{Liu12}. Thirdly, no islands can form in VMEC as a result of the RMPs being applied. There are three dimensional equilibrium codes that can include magnetic islands such as SPEC \cite{Hudson12},  SIESTA \cite{Hirshman11} or HINT2 \cite{Suzuki06}. We hope to investigate the use of these codes for MAST in future work.       
       
\ack

We thank S P Hirshman for the use of VMEC and R Sanchez for the use of COBRA. We also thank T M Bird, W A Cooper, J Harrison, P Helander, S Lazerson, C Nuehrenberg and A Thornton for useful discussions.

This work was funded by the RCUK Energy Programme [grant number EP/I501045] and the European Communities under the contract of Association between EURATOM and CCFE.  To obtain further information on the data and models underlying this paper please contact PublicationsManager@ccfe.ac.uk. The views and opinions expressed herein do not necessarily reflect those of the European Commission.

\section*{References}

\bibliography{VMECbib}{}

\begin{thebibliography}{10}

\bibitem{Loarte03}
Loarte A {\it et al.} (2003) {\em Plasma Phys. Control. Fusion} {\bf 45} 1549

\bibitem{Kirk13}
Kirk A {\it et al.} (2013) {\em Nucl. Fusion} {\bf 53} 043007

\bibitem{Chapman12}
Chapman I T {\it et al.} (2012) {\em Plasma Phys. Control. Fusion} {\bf 54} 105013

\bibitem{Chapman07}
Chapman I T {\it et al.} (2007) {\em Nucl. Fusion} {\bf 47} L36

\bibitem{Canik12}
Canik J {\it et al.} (2012) {\em Nucl. Fusion} {\bf 52} 054004

\bibitem{Fischer12}
Fischer R {\it et al.} (2012) {\em Plasma Phys. Control. Fusion} {\bf 54} 115008

\bibitem{Chapman13}
Chapman I T {\it et al.} (2013) {\em submitted Plasma Phys. Control. Fusion}

\bibitem{Snyder09}
Snyder P B {\it et al.} (2009) {\em Phys. Plasmas} {\bf 16} 056118

\bibitem{Dickinson11}
Dickinson D {\it et al.} (2011) {\em Plasma Phys. Control. Fusion} {\bf 53} 115010


\bibitem{Turnbull13}
Turnbull A {\it et al.} (2013) {\em Phys. Plasmas} {\bf 20} 056114

\bibitem{Liu00}
Liu Y Q {\it et al.} (2000) {\em Phys. Plasmas} {\bf 7} 3681

\bibitem{Liu12}
Liu Y Q {\it et al.} (2012) {\em Plasma Phys. Control. Fusion} {\bf 54} 124013

\bibitem{Ferraro10}
Ferraro N M {\it et al.} (2010) {\em Phys. Plasmas} {\bf 17} 102508

\bibitem{Glasser99}
Glasser A H {\it et al.} (1999) {\em Plasma Phys. Control. Fusion} {\bf 41} A747

\bibitem{Park07}
Park J K {\it et al.} (2007) {\em Phys. Plasmas} {\bf 14} 052110

\bibitem{Hirshman83}
Hirshman S P and Whitson J C (1983) {\em Phys. Fluids} {\bf 26} 3553

\bibitem{Hirshman86}
Hirshman S P {\it et al.} (1986) {\em Compu. Phys. Commun.} {\bf 43} 143

\bibitem{Hudson12}
Hudson S {\it et al.} (2012) {\em Phys. Plasmas} {\bf 19} 112502

\bibitem{Hirshman11}
Hirshman S P {\it et al.} (2011) {\em Phys. Plasmas} {\bf 18} 062504

\bibitem{Suzuki06}
Suzuki Y {\it et al.} (2006) {\em Nucl. Fusion} {\bf 46} L19

\bibitem{Sanchez00}
Sanchez R {\it et al.} (2000) {\em J. Comput. Phys.} {\bf 161} 576

\bibitem{Debock12}
De~Bock M F M {\it et al.} (2012) {\em Plasma Phys. Control. Fusion} {\bf 54} 025001

\bibitem{Harrison13}
Harrison J R {\it et al.}(2013) {\em Nucl. Fusion} submitted

\bibitem{Connor98}
Connor J W {\it et al.} (1998) {\em Phys. Plasmas} {\bf 5} 2687

\bibitem{Sanchez01}
Sanchez R {\it et al.} (2001) {\em Comput. Phys. Commun.} {\bf 135} 82

\bibitem{Connor79}
Connor J W {\it et al.} (1979) {\em Proc. R. Soc. Lond. A.} {\bf 365} 1

\bibitem{Dewar83}
Dewar R L and Glasser A H (1983) {\em Phys. Fluids}, {\bf 26} 3038

\bibitem{Sanchez97}
Sanchez R {\it et al.} (1997) {\em Nucl. Fusion} {\bf 37} 1363

\bibitem{Scannell13}
Scannell R {\it et al.} (2013) {\em Plasma Phys. Control. Fusion} {\bf 55} 035013

\bibitem{Bird13}
Bird T M and Hegna C C (2013) {\em Nucl. Fusion} {\bf 53} 013004

\bibitem{Chapman13b}
Chapman I T {\it et al.} (2013) {\em Phys. Plasmas} {\bf 20} 056101


\end{thebibliography}
\bibliographystyle{unsrt}
\end{document}